\documentclass[aps,prl,twocolumn,amsmath,amssymb,
superscriptaddress,citeautoscript,floatfix]{revtex4}%

\usepackage{graphicx}%
\usepackage{dcolumn}
\usepackage{amsmath}
\usepackage{amssymb}
\usepackage{color}
\usepackage{multirow}

\newcommand{\mSR}{$\mu$SR}
\newcommand{\lem}{LE-$\mu$SR}
\newcommand{\PSI}{Paul Scherrer Institute}

\newcommand{\TCEuS}{$T_{\rm{C}}^{\rm{EuS}}$}

\newcommand{\EF}{$E_{\rm F}$}
\newcommand{\refsubfig}[2]{\hyperref[#1]{\ref*{#1}#2}}

\usepackage{siunitx}
\usepackage{verbatim} 

\usepackage{placeins} 
\usepackage{float}

\usepackage{textcomp} 
\usepackage[hidelinks]{hyperref}
\hypersetup{
colorlinks=true,    
linkcolor=blue,
citecolor=blue,
urlcolor=blue,
pdfborder={0 0 0},
bookmarksopen=true,
bookmarksopenlevel=2,
pdfpagemode=UseOutlines,
pdfstartview={Fit},
pdftitle={Do topology and ferromagnetism cooperate at the EuS/Bi2Se3 
interface?},
pdfauthor={J. A. Krieger},
pdfsubject={Local magnetic and electronic properties at the 
ferromagnet/topological insulator EuS/Bi2Se3 interface.},
pdfkeywords={low energy muons, SX-ARPES, topological insulator, magnetic
insulator},
pdfhighlight= /I
}

\begin{document}

\title {Do topology and ferromagnetism cooperate at the EuS/Bi$_2$Se$_3$ 
interface?}

\author{J.~A.~Krieger}
\email{jonas.krieger@psi.ch}
\affiliation{Laboratory for Muon Spin Spectroscopy, Paul Scherrer
Institute, CH-5232 Villigen PSI, Switzerland}
\affiliation{Swiss Light Source, Paul Scherrer Institute, CH-5232 Villigen
PSI, Switzerland}
\affiliation{Laboratorium f\"ur Festk\"orperphysik,  ETH Z\"urich, CH-8093
Z\"urich, Switzerland}
\author{Y.~Ou}
\affiliation{Francis Bitter Magnet Lab, Massachusetts Institute of 
Technology, Cambridge, Massachusetts 02139, USA}
\author{M.~Caputo}
\affiliation{Swiss Light Source, Paul Scherrer Institute, CH-5232 Villigen
PSI, Switzerland}
\author{A.~Chikina}
\affiliation{Swiss Light Source, Paul Scherrer Institute, CH-5232 Villigen
PSI, Switzerland}
\author{M.~D\"obeli}
\affiliation{Ion Beam Physics, ETH Z\"urich, Otto-Stern-Weg 5, CH-8093
Z\"urich, Switzerland}
\author{M.-A.~Husanu}
\affiliation{Swiss Light Source, Paul Scherrer Institute, CH-5232 Villigen
PSI, Switzerland}
\affiliation{National Institute of Materials Physics, Atomistilor 405A,
077125 Magurele, Romania}
\author{I.~Keren} 
\affiliation{Laboratory for Muon Spin Spectroscopy, Paul Scherrer
Institute, CH-5232 Villigen PSI, Switzerland}
\author{T.~Prokscha}
\affiliation{Laboratory for Muon Spin Spectroscopy, Paul Scherrer
Institute, CH-5232 Villigen PSI, Switzerland}
\author{A.~Suter}
\affiliation{Laboratory for Muon Spin Spectroscopy, Paul Scherrer
Institute, CH-5232 Villigen PSI, Switzerland}
\author{C.-Z.~Chang}
\affiliation{Francis Bitter Magnet Lab, Massachusetts Institute of 
Technology, Cambridge, Massachusetts 02139, USA}
\affiliation{Department of Physics, The Penn State
University, State College, Pennsylvania 16802, USA}
\author{J.~S.~Moodera}
\affiliation{Francis Bitter Magnet Lab, Massachusetts Institute of 
Technology, Cambridge, Massachusetts 02139, USA}
\affiliation{Department of Physics, Massachusetts Institute of Technology,
Cambridge, Massachusetts 02139, USA}
\author{V.~N.~Strocov}
\email{vladimir.strocov@psi.ch}
\affiliation{Swiss Light Source, Paul Scherrer Institute, CH-5232 Villigen
PSI, Switzerland}
\author{Z.~Salman}
\email{zaher.salman@psi.ch}
\affiliation{Laboratory for Muon Spin Spectroscopy, Paul Scherrer
Institute, CH-5232 Villigen PSI, Switzerland}

\date{\today}

\begin{abstract}
  We probe the local magnetic properties of interfaces between the
  insulating ferromagnet EuS and the topological insulator
  Bi$_2$Se$_3$ using low energy muon spin rotation (\lem).  We compare
  these to the interface between EuS and the topologically trivial 
  metal, titanium. Below the magnetic transition of EuS, we detect strong
  local magnetic fields which extend several nm into the adjacent
  layer and cause a complete depolarization of the muons.  However, in
  both Bi$_2$Se$_3$ and titanium we measure similar local magnetic
  fields, implying that their origin is mostly independent of the
  topological properties of the interface electronic states.  In
  addition, we use resonant soft X-ray angle resolved photoemission 
  spectroscopy (SX-ARPES) to probe the electronic band structure at the 
  interface between EuS and Bi$_2$Se$_3$. By tuning the photon energy to the 
  Eu anti-resonance at the Eu $M_5$ pre-edge we are able to detect the 
  Bi$_2$Se$_3$ conduction band,
  through a protective Al$_2$O$_3$ capping layer and the EuS
  layer.  Moreover, we observe a signature of an interface-induced 
  modification of the buried Bi$_2$Se$_3$ wave functions and/or the 
  presence of interface states.
\end{abstract}

\maketitle

\section{Introduction}\label{sec:Intro}

Breaking the time reversal symmetry in topological insulators (TI) 
opens a gap in the topological surface states (TSS)
which are otherwise protected against local perturbations. This has 
been proposed as a route towards several new quantum 
phenomena, such as the
quantum anomalous Hall (QAH) effect~\cite{Yu2010}, the topological 
magneto-electric effect~\cite{Qi2008} and even Majorana excitations, when in 
proximity to an $s$-wave superconductor~\cite{Qi2010}.
The experimental realization of those remains elusive, expect for the QAH 
effect, which exhibits spin polarized,
dissipationless, chiral edge-state transport in the absence of 
external magnetic fields and which has been
observed in  charge compensated, Cr and/or V doped 
TIs~\cite{chang2013,chang2015,Ou2017}.
However, doped TIs suffer from several disadvantages including an
inhomogeneous magnetic gap opening across the surface, partial
magnetic volume fraction at low doping levels and the presence of
impurity bands that can significantly limit their
applicability~\cite{Lee2015,Grauer2015,Lachman2015,Sessi2016,Krieger2017}.
Therefore, the proximity to an insulating magnetic layer was proposed
as an alternative approach to breaking time reversal symmetry at the
surface of a TI.  As a consequence, interfaces between TIs and
magnetic insulators have been investigated with a large number of
different material
combinations~\cite{Wei2013,Yang2013,Lang2014,Assaf2015,%
Katmis2016,Lee2016,Li2015,Huang2017,Li2017,He2017,Tang2017}. 
The hope is that such
interfaces allow for more homogeneous properties across the surface
and induce a magnetic gap via magnetic exchange coupling in the
TSS that forms at the boundary between the TI and the
topologically trivial magnetic insulator.  Another advantage is that
the magnetic transition temperature is given by the choice of the
magnetic layer and can be much higher than for magnetically doped
TIs~\cite{Lang2014,Huang2017,He2017,Tang2017}.
A related promising strategy, proposes to use 
magnetic layers that are chemically similar to the TI and grown directly on 
its surface. This approach, called magnetic extension, has 
recently been 
explored with Bi$_2$MnSe$_4$ based 
compounds~\cite{Hirahara2017,Otrokov2017,Otrokov2017b}.

One of the candidate insulating magnets that has a structure
compatible with the Bi$_2$Se$_3$ TI family is EuS.  The EuS layer
orders ferromagntically in-plane with a Curie temperature \TCEuS\
$\approx\SI{16}{K}$. 
It has been shown with polarized neutron reflectometry that at low 
temperature there is a large induced in-plane magnetic moments 
extending typically $\sim\SI{2}{nm}$  into the 
TI~\cite{Li2015,Katmis2016,Li2017}.  Theoretically,  such an 
in-plane magnetic anisotropy could be sufficient to realize the QAH effect if it 
breaks the reflection symmetry of the TI~\cite{Liu2013}. However, in 
EuS/Bi$_2$Se$_3$ there is evidence for a tilting of the moments at the 
interface, generating an out-of-plane component which can induce a conventional 
exchange gap~\cite{Wei2013,Lee2016}. 
But most surprisingly, it has been reported that a magnetic moment at the 
interface persists up to room temperature~(RT), thereby largely exceeding 
\TCEuS, which makes this interface potentially interesting for spintronics 
application~\cite{Katmis2016}.

The origin of these unusual properties, in particular the high magnetic 
transition temperature, were attributed to the presence of TSS at the 
EuS/TI interface~\cite{Katmis2016,Li2017}.
Indeed, the proximity induced in-plane moment measured in PNR in
charge compensated (Bi,Sb)$_2$Te$_3$/EuS is maximal and decreases
under the application of positive or negative back-gate
voltage~\cite{Li2017}.  This hints at the involvement of the TSS
in the magnetic coupling, but could also be explained by different
screening behaviors of TSS and bulk bands~\cite{Li2017}.

The absence of the QAH effect in current EuS/(charge compensated TI)
devices may be due to a small overlap 
between the TSS and the localized Eu 4f
states which would result in a small 
exchange interaction between the TI and EuS~\cite{Luo2013}.
Moreover, the exchange coupling should be
significant only on a length scale of a few \si{\angstrom} and the 
formation of a topologically trivial interface state is
expected~\cite{Menshov2013}.  Density functional theory (DFT)
calculations on EuS/Bi$_2$Se$_3$ confirm the formation of 
such a trivial interface state inside the bandgap of the TI and
suggest that the topological state is almost gapless for thick
Bi$_2$Se$_3$ layers~\cite{Lee2014,Eremeev2015,Kim2017}.  This is
attributed to the fact that the TSS are shifted away form the
interface and deeper into the TI~\cite{Eremeev2015}. 
Experimentally, the absence of EuS's Raman peaks in the presence of a adjacent 
Bi$_2$Se$_3$ layer points to the presence of significant band-bending in 
EuS~\cite{Osterhoudt2018}. Therefore, the nature of the magnetism at the 
EuS/Bi$_2$Se$_3$ interface remains unclear and highly debated, in particular 
with regards to the interplay between topology and magnetism.

Here, we address this question directly using depth resolved measurements 
of the local magnetic and
electronic properties of EuS/Bi$_2$Se$_3$ heterostructures using muon spin 
spectroscopy (\mSR) and soft X-ray angle resolved photoemission spectroscopy 
(SX-ARPES) at the buried interface.  By tuning the photon energy ($h\nu$) to the 
Eu anti-resonance at the Eu $M_5$ pre-edge we find a clear photoemission signal 
of the Bi$_2$Se$_3$ conduction band, through a protective Al$_2$O$_3$ capping
layer and the EuS layer.  This allows us to confirm that the
electronic structure of the buried Bi$_2$Se$_3$ layer is preserved in
the presence of the EuS and capping layer.  Our \mSR\ measurements show 
that below the magnetic transition of EuS, there are strong local magnetic
fields which extend several nanometers into the adjacent TI layer and
completely depolarize the muons.  Comparison between the properties of
the EuS/Bi$_2$Se$_3$ and EuS/titanium interfaces reveals that they
are very similar magnetically, implying that the presence of TSS
at the interface are most probably not a dominant factor in the observed 
proximity effect at these interfaces.

\section{Experiment}\label{sec:Exp}

The studied samples consist of layers of Bi$_2$Se$_3$,
V$_{0.2}$(Bi$_{0.32}$Sb$_{0.68}$)$_{1.8}$Te$_{3}$ and Ti grown onto
sapphire (0001)~substrates by molecular beam
epitaxy~\cite{Zhang2011,chang2013b}. A layer of EuS was added by
evaporation using an electron-beam
source at room temperature~\cite{Katmis2016}. Finally, all samples 
were capped with an amorphous Al$_2$O$_3$ layer to protect them during 
\textit{ex-situ} transportation. 
The thickness of both the Al$_2$O$_3$ and EuS was  
$\SI{4}{nm}$ for the \mSR\ experiments and \SI{1}{nm} for ARPES.
All samples and their corresponding layer compositions are listed 
in Table~\ref{tab:Samples}.
\begin{table}[ht]
  \resizebox{\columnwidth}{!}{
    \centering
    \begin{tabular}{lcll}\hline\hline
     \multicolumn{1}{c}{Cap}&EuS& \multicolumn{1}{c}{Interlayer}&
     \multicolumn{1}{c}{Technique}\\ \hline
      \SI{4}{nm}\,Al$_2$O$_3$ &~\SI{4}{nm}~ &20\,QL~Bi$_2$Se$_3$ &   \lem\\
      \SI{4}{nm}\,Al$_2$O$_3$ 
&\SI{4}{nm}&20\,QL~V$_{0.2}$(Bi$_{0.32}$Sb$_{0.68}$)$_{1.8}$Te$_3$ &   \lem\\
      \SI{4}{nm}\,Al$_2$O$_3$ &\SI{4}{nm} &\SI{60}{nm}~Ti & \lem\\
      \SI{4}{nm}\,Al$_2$O$_3$ &\SI{4}{nm}&60\,QL~Bi$_2$Se$_3$ &   \lem\\
      \SI{1}{nm}\,Al$_2$O$_3$ &\SI{1}{nm}&10\,QL~Bi$_2$Se$_3$ &   SX-ARPES\\
      \SI{10}{nm}\,Se &-&10\,QL~Bi$_2$Se$_3$ &   SX-ARPES\\
     \hline\hline
     \end{tabular}}
     \caption{ Nominal thicknesses of the investigated samples, grown on 
sapphire~(0001) substrates.}\label{tab:Samples}
\end{table}
The thickness of the topological insulators is given in quintuple layers 
(\SI{1}{QL}~$\approx\SI{1}{nm}$).
The layer thickness and interface
quality of the \mSR\ samples has been verified by Rutherford
backscattering (RBS) at the Tandem accelerator of ETH Zurich.

The SX-ARPES experiments were performed with p-polarized
light on the ADRESS beamline (X03MA) at the Swiss Light Source, Paul
Scherrer Institut, Villigen, Switzerland~\cite{Strocov2010}. 
During the measurements the temperature was kept below \SI{12}{K} and the 
analyzer slit was oriented along the incident X-ray 
direction~\cite{Strocov2014}.
The combined beamline and analyzer
resolution at $h\nu=\SI{1.12}{keV}$ was better than
\SI{220}{meV}. 
The heterostructures were
probed through the amorphous \SI{1}{nm} Al$_2$O$_3$ capping layer.  The
higher photoelectron escape depth of SX-ARPES in comparison to
standard UV-ARPES allows to retrieve information from underneath such
a thin layer~\cite{Kobayashi2012}.    In addition, we investigated
reference samples of Bi$_2$Se$_3$ protected by a Se capping layer,
which was removed \textit{in situ} before the
measurement~\cite{Hoefer2015}. 
All samples were investigated with the same beamline and 
analyzer settings.
Supporting X-ray absorption spectra (XAS) were
recorded \textit{in-situ} by detecting the total electron yield (TEY)
via the drain current of the sample.

The low energy \mSR\ experiments were performed on the $\mu$E4
beamline of the Swiss Muon Source at \PSI\ in Villigen,
Switzerland~\cite{Prokscha2008}. Fully spin-polarized muons were implanted into 
the sample with an implantation energy, $E$, tunable from \SI{1}{keV} to
\SI{12}{keV}. The muons decay with a lifetime of
$\approx\SI{2.2}{\micro s}$ into a positron and two neutrinos.  Parity
violation of this weak decay dictates that the decay positron is
emitted preferentially along the muon spin
direction~\cite{Garwin1957}.  Therefore, measuring the spatial
distribution of the decay positrons with four detectors around the
sample allows us to determine the ensemble average of the temporal
evolution of the muon spin polarization.  For these measurements the
samples were glued on a Ni-coated sample plate, which suppresses the
signal from muons missing the sample~\cite{Saadaoui2012}. The
measurements were performed in the temperature range of
\SIrange{4}{320}{K} and in a weak transverse field (wTF) of
\SI{5}{mT}, which was applied perpendicular to the sample surface. 
The data was analyzed with the \texttt{Musrfit}
software~\cite{Suter2012}.
The muon stopping distributions as a function of energy were modeled
with the \texttt{Trim.SP} code~\cite{morenzoni2002}.

\section{Results}\label{sec:Results}
\subsection{Structural characterization using RBS}
\begin{figure}[htb]
  \includegraphics[width=0.92\linewidth]{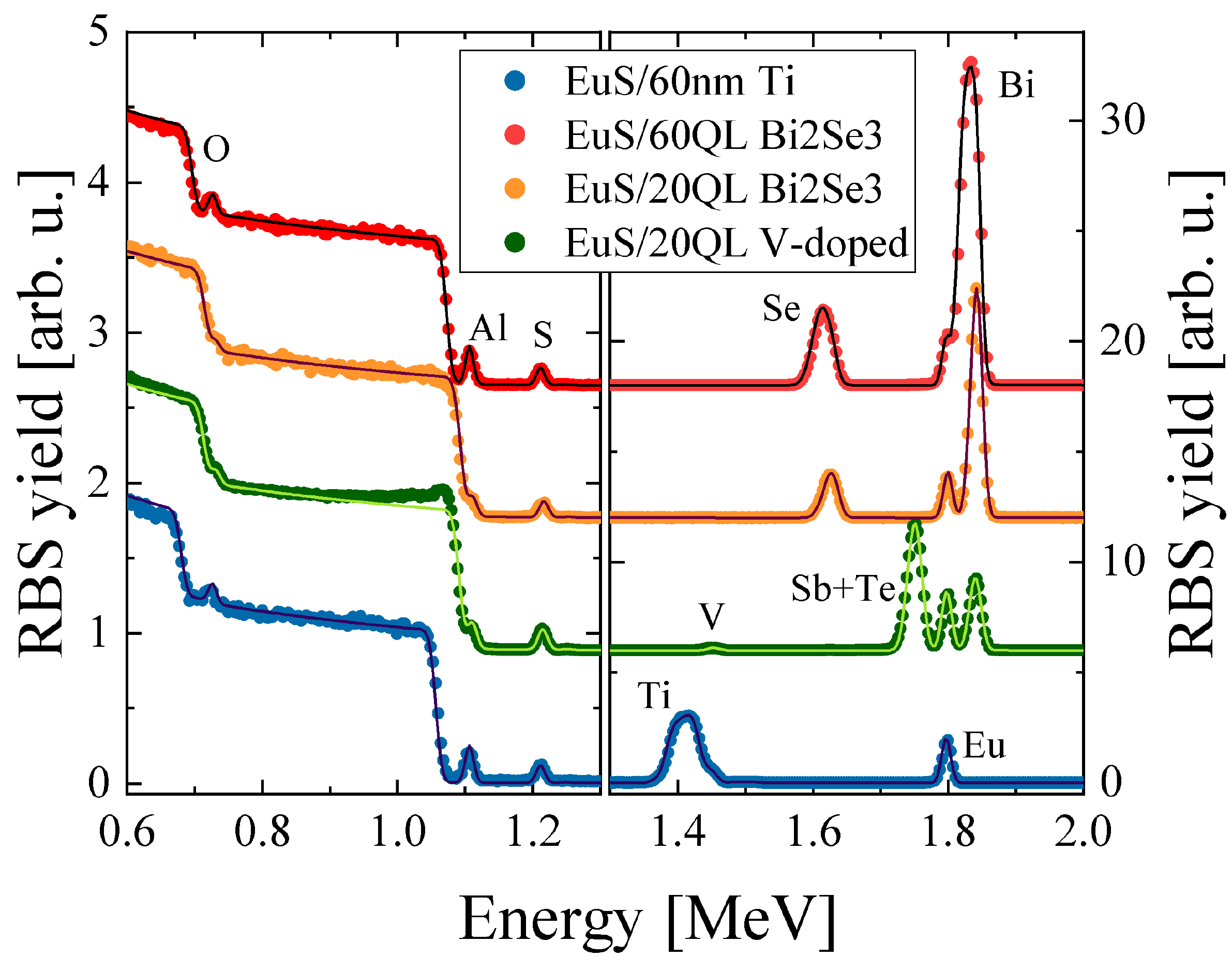}
  \caption{RBS spectrum of \SI{2}{MeV} He ions at a backscattering angle
    of $168^\circ$. The data is presented using a different offset
    and axis-scale in the left and right panel.}
\label{fig:RBS}
\end{figure}

The thickness and stoichiometric properties of the layers were
verified using RBS measurements.  The RBS yield as a function of final
He energy is shown in Fig.~\ref{fig:RBS}.
The resulting layer thicknesses from these measurements are given in
Table~\ref{tab:thicknesses}.
\begin{table}[ht]
  \resizebox{\columnwidth}{!}{
    \centering
    \begin{tabular}{lccc}\hline\hline
     \multicolumn{1}{c}{Interlayer}  &$\rho_{\mathrm{Al_2O_3}}$
&$\rho_{\mathrm{EuS}}$& $\rho_{\mathrm{Interlayer}}$\\ 
\hline
20\,QL~Bi$_2$Se$_3$ &\SI{32.5}{} &\SI{16.7}{} &\SI{83.3}{}  \\
20\,QL~V$_{0.2}$(Bi$_{0.32}$Sb$_{0.68}$)$_{1.8}$Te$_3$ & 
\SI{47.5}{}&\SI{22.8}{}&\SI{74}{} \\
60\,QL~Bi$_2$Se$_3$ & \SI{55}{}&\SI{15.1}{}&\SI{178.9}{} \\
\SI{60}{nm}~Ti &\SI{55}{}&\SI{15.2}{}&-\\
     \hline\hline
     \end{tabular}}
     \caption{Area number density of the layers  determined 
by 
RBS in $10^{15}$~at./cm$^2$ of the samples investigated by 
\lem. }\label{tab:thicknesses}
\end{table}
The listed values were used as input parameter for 
all subsequent analysis. The composition of the various layers in
the studied samples are confirmed to be free of impurities, except for
the Ti layer which contains some additional transition metals (less
than \SI{20}{at{.}\percent} of mostly V and Co).  In all samples, the
EuS layer is found to be slightly S deficient, with the ratio Eu/S
ranging from $0.85$ to $0.96(3)$. The samples with \SI{60}{nm} thick
interlayers exhibit a sharp EuS/interlayer interface, whereas in the
\SI{20}{nm} samples the interlayer is extending slightly into the EuS
layer. This could be due to either interface roughness or intermixing,
which cannot be distinguished by RBS.

\subsection{Electronic Properties using SX-ARPES}

\begin{figure*}[htb]
  \includegraphics[width=.999\linewidth]{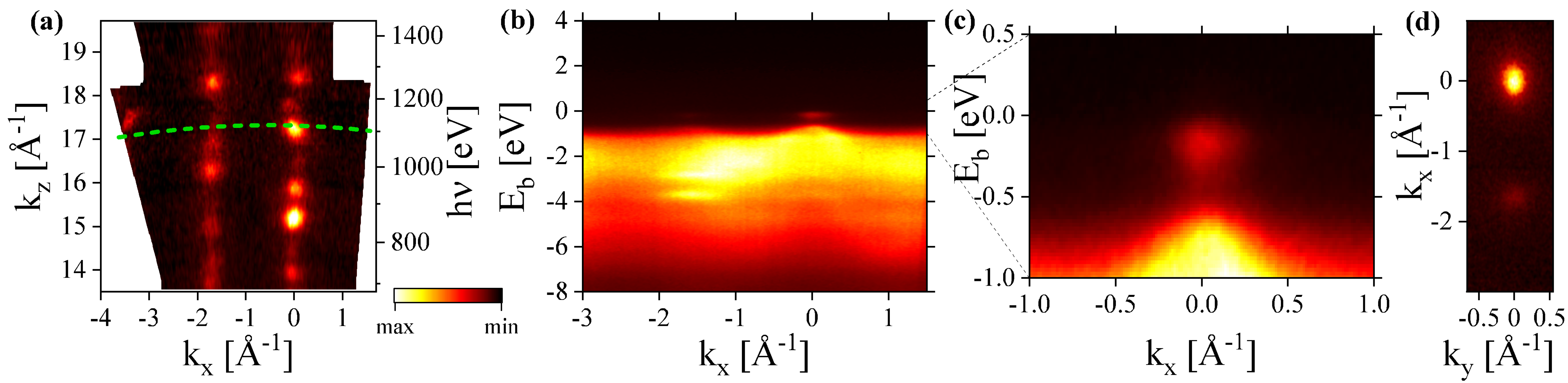}
 \caption{ \textbf{(a)} Fermi surface map of bare Bi$_2$Se$_3$ along
   ($k_x$,~$k_z$). The corresponding $h\nu$ values at $k_x=0$ are 
shown in the right
   axis. The dashed line indicates $h\nu=\SI{1120}{eV}$.
   \textbf{(b),(c)} High statistics cut at $h\nu=\SI{1120}{eV}$
   showing the conduction band and hints of the surface states of
   Bi$_2$Se$_3$.
   \textbf{(d)} Fermi surface cut around the $\Gamma$ point at 
$h\nu=\SI{1120}{eV}$.}
\label{fig:Bi2Se3}
\end{figure*}

The ARPES intensity form a buried layer is usually very
small.
It is therefore helpful to first characterize a reference Bi$_2$Se$_3$ 
sample independently before considering the full heterostructure.  In
Figure~\refsubfig{fig:Bi2Se3}{(a)} we show  the out-of-plane momentum 
$k_z$ dependence (rendered from h$\nu$) of the ARPES intensity of bare 
Bi$_2$Se$_3$ at the Fermi level (\EF) along the $\Gamma$-M direction.
The observed  Fermi intensity, composed of contributions from the conduction 
band and the TSS, exhibits periodic oscillations across the different $\Gamma$ 
points in $k_z$, where the relative weight of the two components can 
vary~\cite{Queiroz2016}.
A representative photoemission 
spectrum and a 
Fermi surface measured at $h\nu=\SI{1120}{eV}$ are shown in  
Figs.~\refsubfig{fig:Bi2Se3}{(b,c)} and Fig.~\refsubfig{fig:Bi2Se3}{(d)}, 
respectively.

\begin{figure}[htb]
  \includegraphics[width=.95\linewidth]{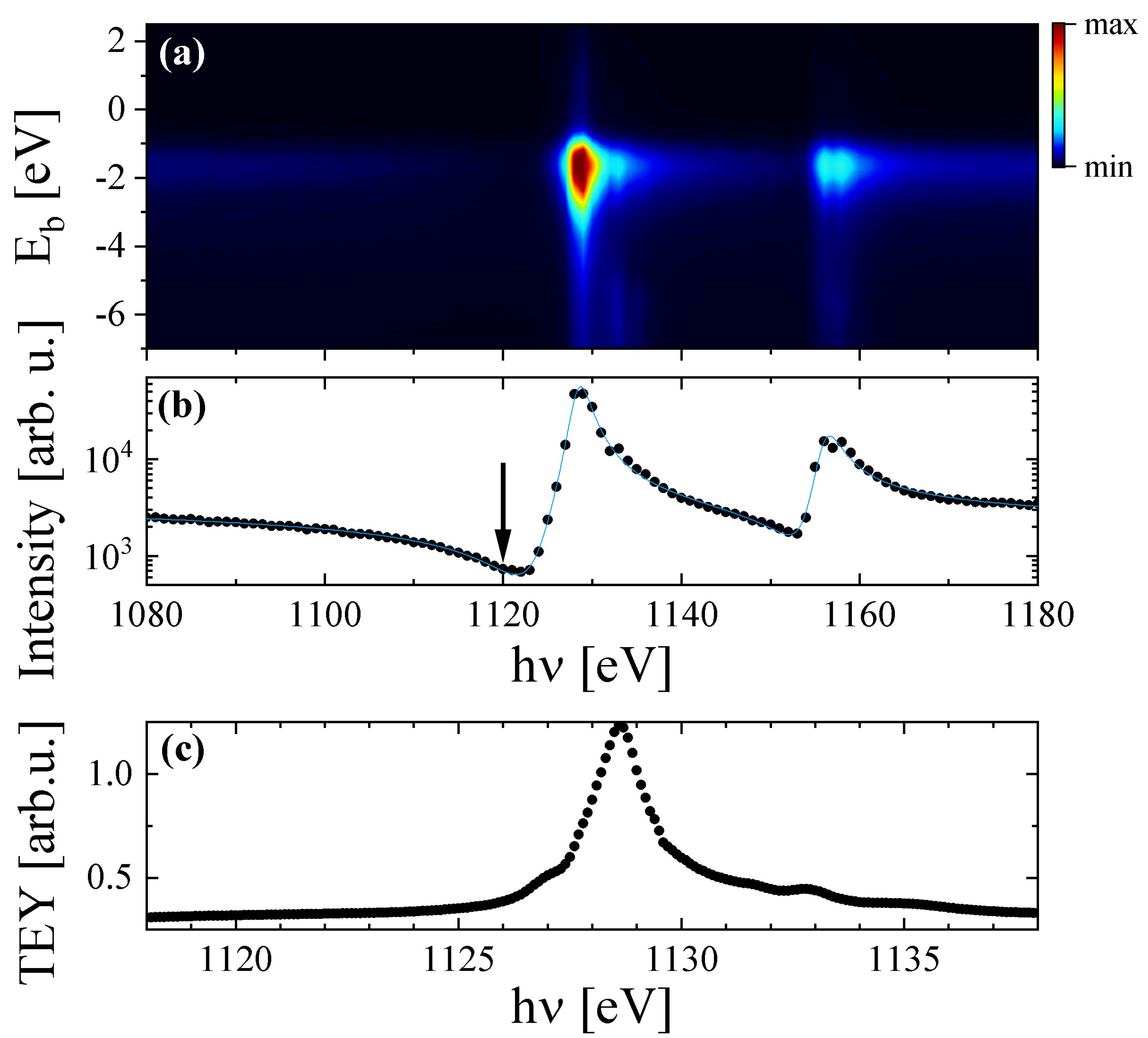}
  \caption{\textbf{(a)} Angular integrated PES showing the resonating
    valence states at the Eu M$_4$ and M$_5$ edges. \textbf{(b)}
    Intensity of the resonating Eu peak (at E$_b\approx\SI{-1.7}{eV}$). The 
    solid line depicts a sum of two Fano profiles and a linear
    background. The arrow indicates $h\nu=\SI{1120}{eV}$.
    \textbf{(c)} XAS at the Eu M$_5$ edge.}
\label{fig:ResPES}
\end{figure}

For the heterostructure samples, which have been capped with an 
Al$_2$O$_3$ layer 
of \SI{1}{nm} thickness, we have confirmed the absence of any 
significant degradation by checking the Eu valence using XAS.    
The shape and position of the
 Eu M$_5$ XAS peak in Fig.~\refsubfig{fig:ResPES}{(c)} clearly 
shows that Eu is mostly in the ferromagnetic
Eu$^{2+}$ state, cf.~Ref.~\cite{Thole1985,Lev2016}. However, we note that 
samples
which were stored \textit{ex-situ} (for several weeks) developed a
considerable weight of Eu$^{3+}$. We suspect this is because of
oxidation of the Eu through the thin capping layer. 

Results of resonant photoemission spectroscopy measurements across the 
Eu M$_{4,5}$ edges are shown in Fig.~\refsubfig{fig:ResPES}{(a)}. 
In the vicinity of the Eu M~edge, there is an enhanced cross section for 
coherent photoemission via intermediate 3d$^9$4f$^{n+1}$ 
states, where $n=6$ or $7$ for Eu$^{3+}$ or Eu$^{2+}$, respectively. These 
second-order processes can 
interfere with direct photoemssion, leading to a Fano-like lineshape of the 
intensity as a function of $h\nu$~\cite{Fano1961}.
A comparison to the XAS spectrum reveals that the Eu$^{2+}$ 
resonates around E$_b\approx\SI{-1.7}{eV}$, whereas a small 
resonance of Eu$^{3+}$ atoms is found at higher $h\nu$ around 
E$_b\approx\SI{-5}{eV}$.
Figure~\refsubfig{fig:ResPES}{(b)}
shows the integrated intensity of the Eu$^{2+}$ PES peak across the Eu M$_5$ and 
M$_4$ edges.
As expected, the resonant photoemission intensity follows a Fano-profile 
with a pronounced anti-resonance at the pre-edge. A similar anti-resonance 
behavior is often observed in resonant photoemission on transition 
metals~\cite{Robey1992,Stadnik1994,Weinelt1997}.

\begin{figure}[htb]
 \includegraphics[width=.84\linewidth]{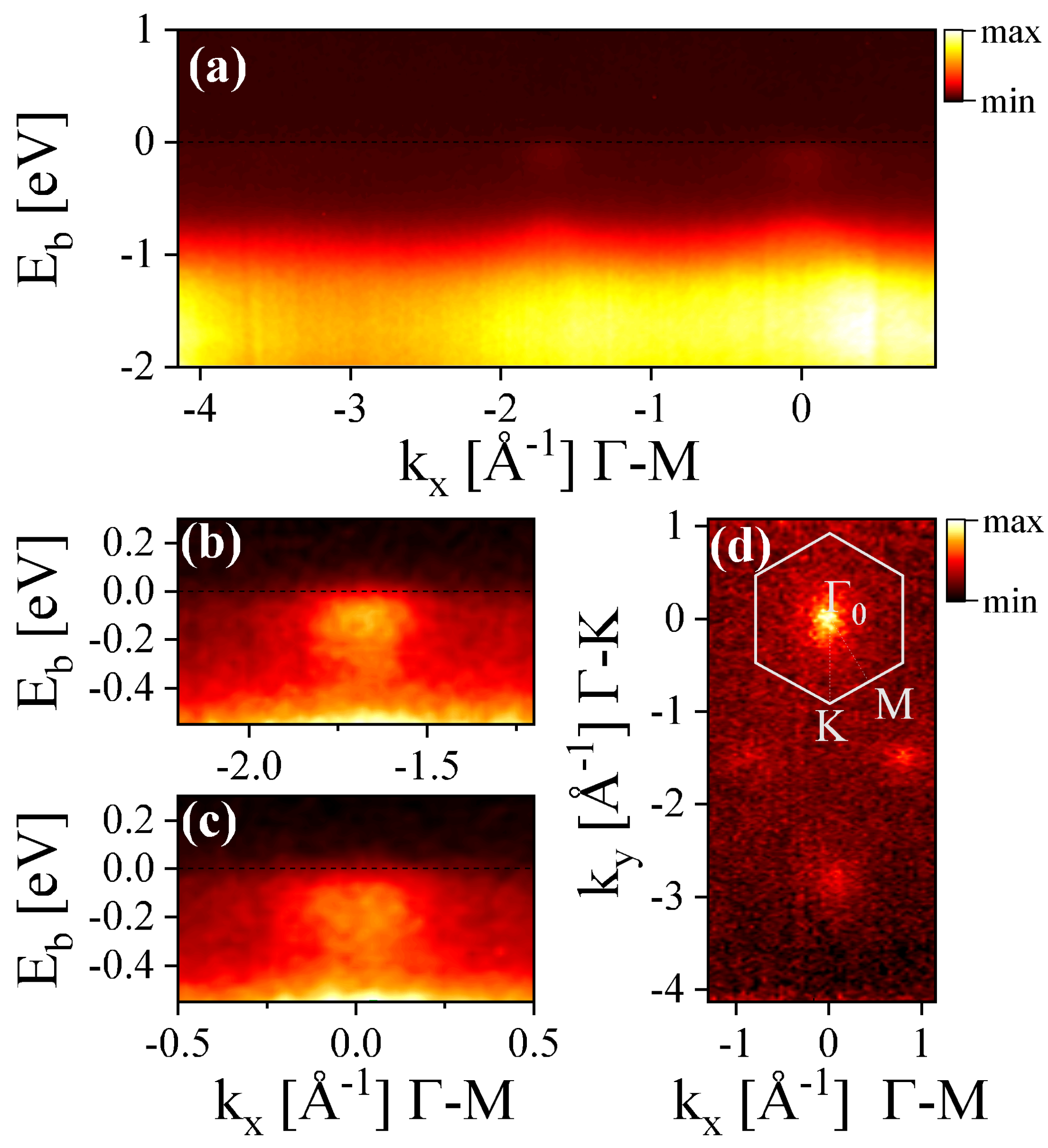}
 \caption{
   \textbf{(a)} Band structure of the buried Bi$_2$Se$_3$ at
   $h\nu=\SI{1120}{eV}$ along $\Gamma$-M and seen though the caps of
   Al$_2$O$_3$ and EuS.
   \textbf{(b),(c)} Zoom on the $\Gamma_1$ and $\Gamma_0$
   points, respectively.  \textbf{(d)} Fermi surface measured with
   the sample oriented along the $\Gamma$-K direction.  }
\label{fig:buriedTI}
\end{figure}

Despite the  large probing depth of SX-ARPES, observation of a weak dispersive
signal from the buried Bi$_2$Se$_3$ is hindered by overwhelming intensity 
around E$_b\approx\SI{-1.7}{eV}$, which mainly corresponds to 
a $^{7}$F final state multiplett excited from 
Eu$^{2+}$~[\onlinecite{Gerrit1993,Yamamoto2005},~%
Fig.\,\refsubfig{fig:ResPES}{(a)}]. 
However, the Eu M$_5$ anti-resonance at 
${h\nu=\SI{1120}{eV}}$ offers a
favorable photon energy to ``see through'' the capping layers:
The size of the Eu contribution around $E_b\approx\SI{-1.7}{eV}$ is
reduced by almost a factor of~4~[Fig.~\ref{fig:ResPES}] and it lies close 
to a $\Gamma$ point of bare Bi$_2$Se$_3$ in $k_z$, where the conduction band and
the surface states are expected to be seen [Fig.~\ref{fig:Bi2Se3}].
A high statistics measurement at this energy is shown in
Fig.~\refsubfig{fig:buriedTI}{(a)-(c)}. It reveals dispersive spectral weight
close to \EF\ at the $\Gamma$ point. 
Since Al$_2$O$_3$ is amorphous, the Eu 4f electrons in EuS are very 
localized and both layers are insulating,  none of them should exhibit a 
dispersion close to \EF.
Therefore, the observed dispersing features come from 
the buried Bi$_2$Se$_3$ layer or its interface with EuS.
To confirm this origin, we show a Fermi surface cut at the 
same h$\nu$ in Fig.~\refsubfig{fig:buriedTI}{(d)}. It exhibits a 
hexagonal Brillouin zone pattern characteristic of  Bi$_2$Se$_3$.

\subsection{Magnetic Properties using LE-$\mathbf{\mu}$SR}

\begin{figure}[htb]
  \includegraphics[width=0.93\linewidth]{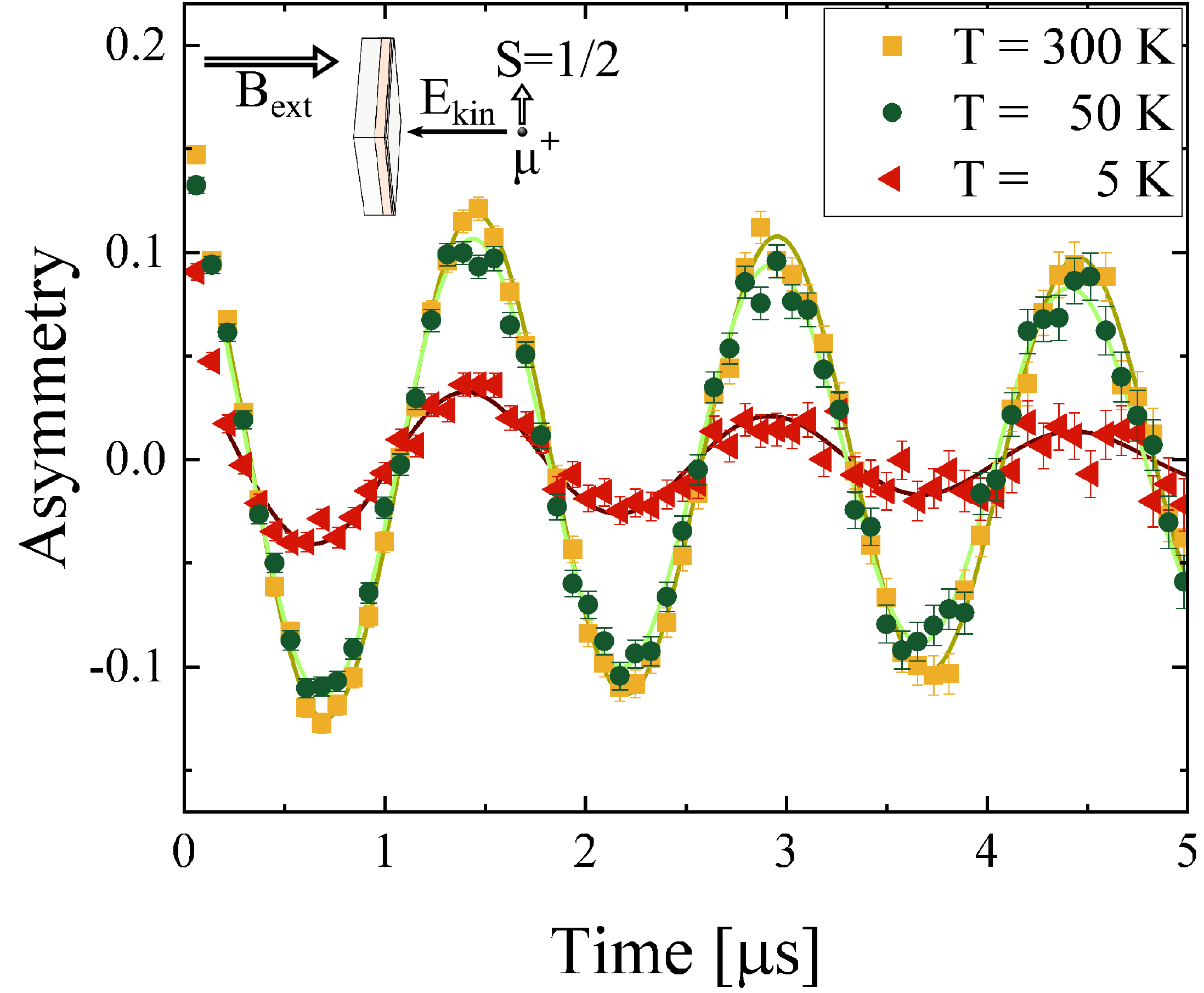}
  \caption{ Weak transverse field asymmetry spectra of \SI{1.5}{keV} 
    muons in the \SI{4}{nm}~Al$_2$O$3$/\SI{4}{nm}~EuS/20\,QL~Bi$_2$Se$_3$ 
    sample at different temperatures in an applied field of \SI{5}{mT}. The 
    inset depicts the measurement geometry.}
\label{fig:spectrum}
\end{figure}

Representative \mSR\ asymmetry spectra of 
EuS/(20\,QL)\,Bi$_2$Se$_3$ 
are shown in Fig.~\ref{fig:spectrum}.
The measured asymmetry exhibits a weakly damped oscillation at room
temperature, as is typical for a paramagnetic 
sample~\cite{yaouanc2011}.
With decreasing temperature, there is a slight reduction of the oscillation 
amplitude and at \SI{5}{K} it becomes a much smaller and the oscillation 
is heavily damped. 
This indicates that the implanted muons experience a broad
distribution of magnetic fields in a part of the sample, 
particularly in the EuS layer.  There is an additional fast depolarization of a
small part of the signal that is attributed to muons stopping in the
magnetic Ni coated sample holder and in the sapphire capping layer and
substrate~\cite{Saadaoui2012,brewer2000}.
These contributions have been subtracted by fitting the data
measured after \SI{0.2}{\micro s} to an exponentially damped
cosine~\cite{Krieger2017}, see appendix for details.

\begin{figure*}[htb]
  \includegraphics[width=.98\linewidth]{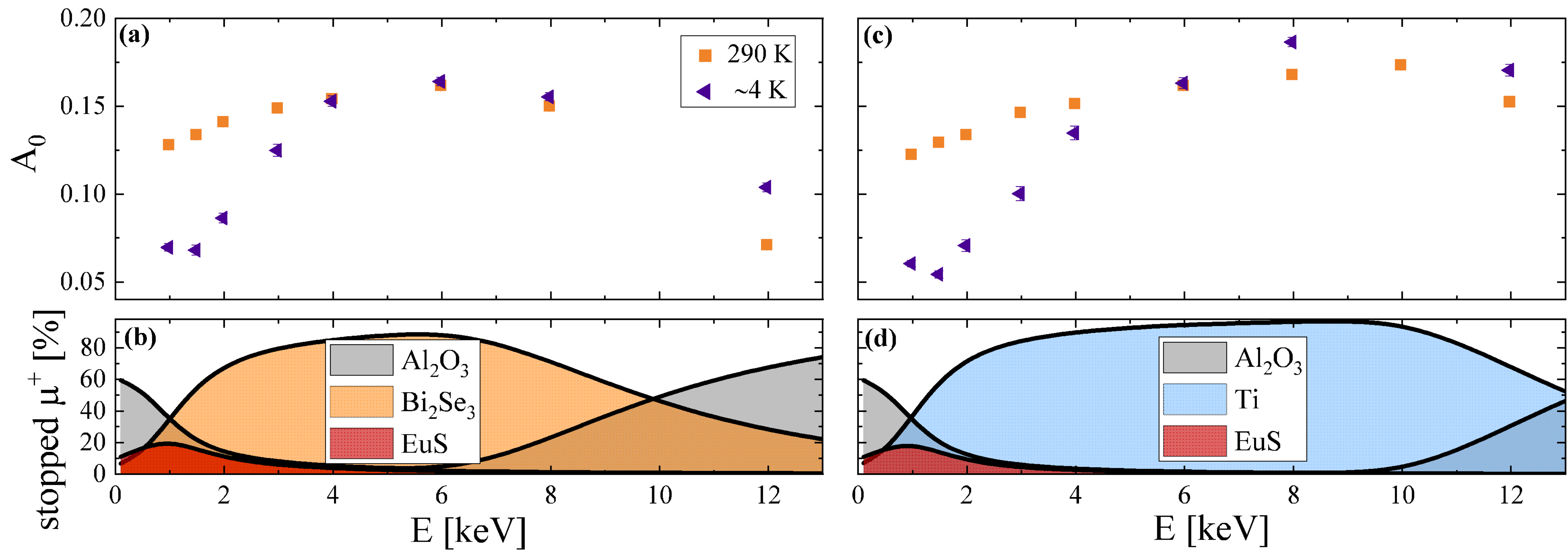}
  \caption{Initial asymmetry $A_0$ as a function of implantation energy $E$ 
    in the~\textbf{(a)} EuS/(60\,QL)\,Bi$_2$Se$_3$  and~\textbf{(c)} 
    EuS/(\SI{60}{nm})Ti samples. 
    \textbf{(b)} and~\textbf{(d)} show the corresponding
    calculated stopping fractions.}
\label{fig:EScans}
\end{figure*}

\begin{figure}[htb]
  \includegraphics[width=0.95\linewidth]{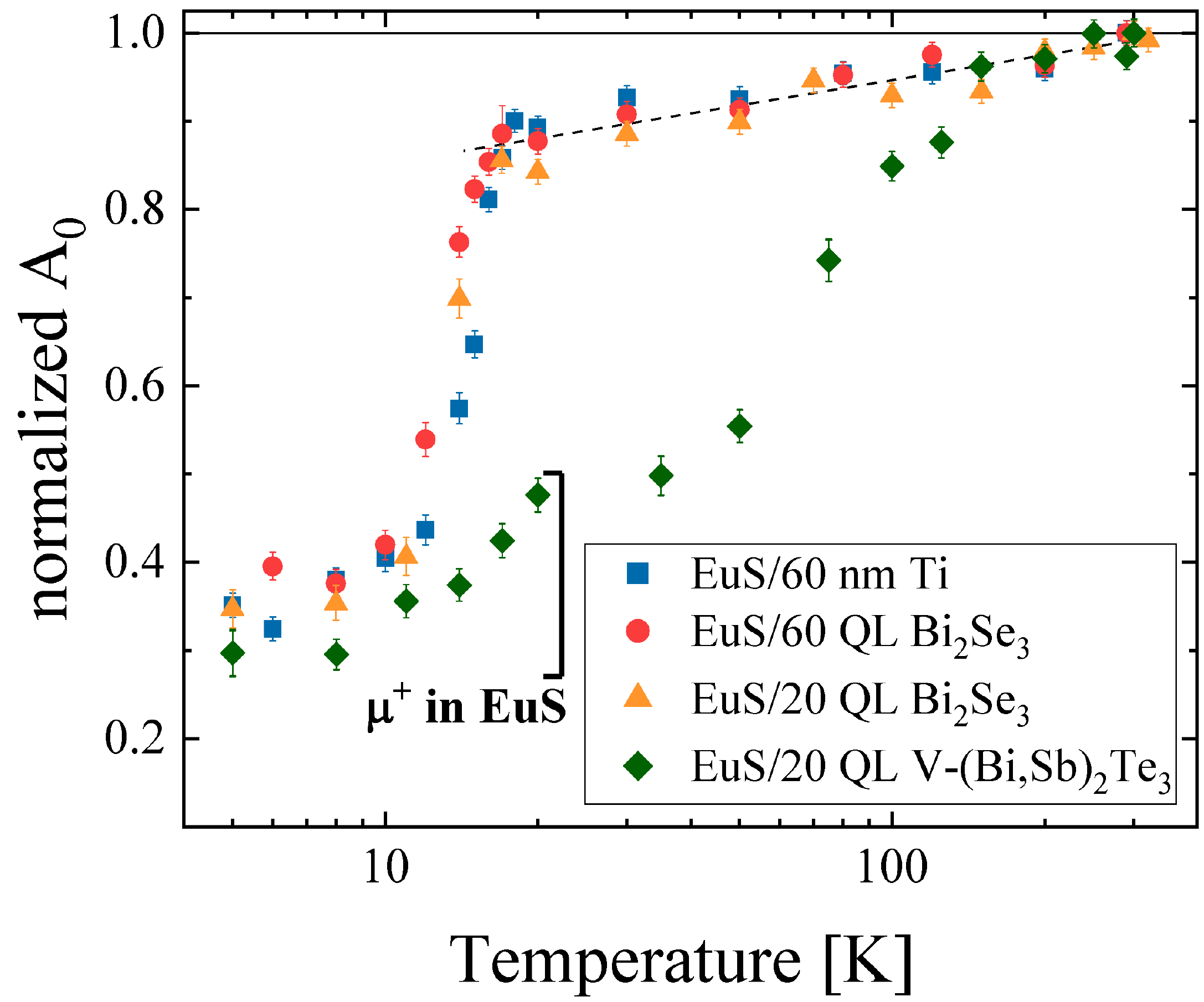}
  \caption{Initial asymmetry  at \SI{1.5}{keV} implantation energy 
normalized to the RT value as a function of
    temperature. The black bar indicates the fraction of muons
    stopping in the EuS layer of the V-doped sample. The black lines are guides 
to the eye.
  }
\label{fig:nAsy}
\end{figure}

The initial asymmetry, $A_0$, as a function of the implantation energy $E$ for 
EuS/(60\,QL)\,Bi$_2$Se$_3$ and EuS/(\SI{60}{nm})Ti  
is shown in
Figs.~\refsubfig{fig:EScans}{(a)} and~\refsubfig{fig:EScans}{(c)}, 
respectively~\footnote{
Note that we suspect that the sample wasn't fully 
thermalized during the measurement shown in Fig.~\refsubfig{fig:EScans}{(a)}. A 
comparison to the temperature dependence in Fig.~\ref{fig:nAsy} gives an 
effective temperature of $\approx\SI{13}{K}$.
}.
We find that the behavior of $A_0$ depends strongly on the implantation energy 
$E$ and thereby on the probed layer, cf. 
Figs.~\refsubfig{fig:EScans}{(b)}~and~\refsubfig{fig:EScans}{(d)}. 
The signal is almost temperature independent at intermediate
$E$ ($\sim 5-\SI{8}{keV}$), where most of the muons stop deep in the 
interlayer, whereas  at low $E$ ($\lesssim\SI{4}{keV}$), where most of the 
muons stop in the vicinity of the EuS/interlayer interface, there is a large 
drop in $A_0$ as the temperature is decreased.

In Figure~\ref{fig:nAsy} we compare the temperature dependence of
this drop for different samples. The data sets have been normalized to their
RT values.  An implantation energy of $E=\SI{1.5}{keV}$ was used, in 
order to maximize the number of muons stopping close 
to the EuS/interlayer interface.
All samples exhibit a gradual decrease in $A_0$ with 
decreasing temperature and a sharp drop below 
\TCEuS$\sim\SI{16}{K}$.
Additional measurements in zero and longitudinal magnetic field (not 
shown)
indicate that this drop is primarily due to static magnetism,
causing an additional depolarization of the muon spin that can be almost fully 
decoupled upon application of \SI{10}{mT} longitudinal field.

The V$_{0.2}$(Bi,Sb)$_{1.8}$Te$_{3}$ layer is expected to
have a broad magnetic transition with an onset around
${T_{\mathrm C}\approx\SI{150}{K}}$~\cite{Krieger2017}. 
Indeed we observe two
sequential drops of $A_0$ in 
EuS/V$_{0.2}$(Bi,Sb)$_{1.8}$Te$_3$\,: One below \SI{150}{K} 
followed by a second one at $\sim\SI{16}{K}$, corresponding to 
\TCEuS~[Fig.~\ref{fig:nAsy}].  The 
first drop is
accompanied by both a decrease of the mean field and an increase of
the depolarization rate [Fig.~\ref{fig:wTF}] which is consistent with
our previous measurements~\cite{Krieger2017}.  Both of these
properties are mostly unaffected by the magnetic transition of the EuS
layer. This is not very surprising, since 
the signal from the V-doped layer is lost already above \TCEuS . 
However, this situation is different in the other
samples, where the transition of EuS is accompanied by a decrease of
the mean field in the sample and a peak in the depolarization rate
[Figs.~\ref{fig:wTF}~and~\ref{fig:Ti}].

\section{Discussion }\label{sec:Disc}
\subsection{Electronic properties}

In DFT calculations of EuS/Bi$_2$Se$_3$, a sharp interface between a Se and Eu 
layer is typically assumed. This leads to the presence of a topologically
trivial interface state that crosses the Fermi surface between
$\Gamma$-K and forms a plateau  at the M
point along $\Gamma$-M around
$E_b\approx\SI{-0.2}{eV}$~\cite{Eremeev2015,Kim2017,Lee2014}. Some calculations
further predict an EuS derived band which dips below \EF\ at the
M-point~\cite{Kim2017}. 
However, recent DFT results suggest that the presence 
of these trivial states depends on the assumed interface 
structure~\cite{Eremeev2018}.
Indeed, we observe none of these bands
experimentally. This may be due to a different arrangement of the atoms 
at the interface than what was originally assumed in DFT. But we cannot 
exclude the presence of interface roughness which could prevent
the interface states from forming with a clear in-plane dispersion,
 or simply a very low photoemission cross section with the interface 
states at $h\nu=\SI{1120}{eV}$. 
Even in bare Bi$_2$Se$_3$ the instrumental
resolution is insufficient to resolve the dispersion of the TSS
and distinguish it from the bulk conduction band,
Fig.~\refsubfig{fig:Bi2Se3}{(c)}.  Hence, it is possible that the spectrum
mainly consists of the bulk conduction band that is smeared across the
gap by the experimental resolution.

\begin{figure}[htb]
 \includegraphics[width=.90\linewidth]{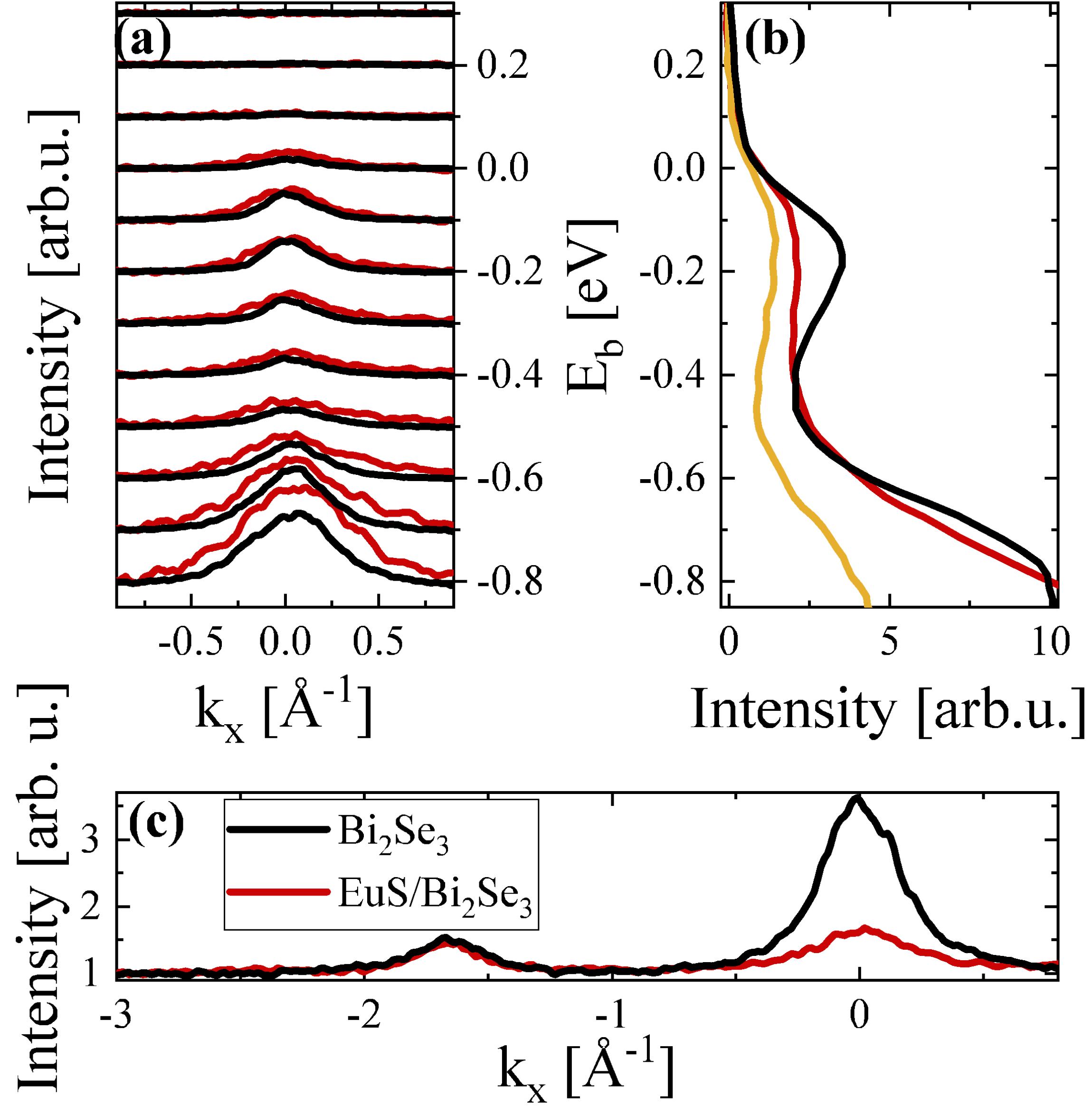}
 \caption{ \textbf{(a)} MDCs of 
 EuS/(10\,QL)\,Bi$_2$Se$_3$ 
  in comparison to MDCs of bare Bi$_2$Se$_3$.  
   \textbf{(b)}  EDC at the $\Gamma$ point 
integrated within $k_x=\pm\SI{0.1}{\angstrom^{-1}}$. In order to reduce the
   contribution form non-dispersive spectral weight in 
EuS/(10\,QL)\,Bi$_2$Se$_3$, the yellow line
   shows the difference between the EDC around $\Gamma$ and around M.
   \textbf{(c)} Total intensity within \SI{-0.3}{eV} below \EF\ and within 
$k_y=\pm\SI{0.15}{\angstrom^{-1}}$ around $\Gamma$. 
   The scale between measured intensities on EuS/(10\,QL)\,Bi$_2$Se$_3$ 
and bare Bi$_2$Se$_3$ is arbitrary and different in each subplot.
 }
\label{fig:Cuts}
\end{figure}

A detailed comparison of the momentum and energy distribution curves (MDC and 
EDC, respectively) of the EuS/Bi$_2$Se$_3$ and
bare Bi$_2$Se$_3$ reveals a very similar 
dispersion~[Fig.~\ref{fig:Cuts}].
The MDCs around 
the $\Gamma_0$ point and within the band gap of EuS are 
only slightly broader in EuS/Bi$_2$Se$_3$ 
compared to Bi$_2$Se$_3$.
In contrast, the EDCs at $\Gamma_0$ are different in the two samples:  
We observe a plateau between \EF\ and \SI{-0.6}{eV} in
EuS/Bi$_2$Se$_3$,
whereas in bare Bi$_2$Se$_3$ there is a clear dip around
\SI{-0.45}{eV}. However, in EuS/Bi$_2$Se$_3$ 
a significant part of the spectral weight is non-dispersive. In order to 
compare only the dispersive part to bare Bi$_2$Se$_3$ we consider the
difference between the EDCs at $\Gamma$ and M, where we
don't see any dispersion. This is shown as a yellow line in 
Fig.~\refsubfig{fig:Cuts}{(b)}, revealing a qualitatively similar behavior to 
bare Bi$_2$Se$_3$. This indicates that the dispersive line shape of the buried 
Bi$_2$Se$_3$ remains mostly unaffected by the presence of EuS.

Nevertheless, we note a clear discrepancy between EuS/Bi$_2$Se$_3$ 
and bare Bi$_2$Se$_3$ in the relative intensity
of the conduction band at the $\Gamma_0$ and $\Gamma_1$ points. While 
the intensity at $\Gamma_1$ is much lower than at $\Gamma_0$
in bare
Bi$_2$Se$_3$, the two points have a comparable spectral weight in
EuS/Bi$_2$Se$_3$~[Figs.~\refsubfig{fig:Bi2Se3}{(b)}~and~%
\refsubfig{fig:buriedTI}{(a)}].
To exclude that this is an artifact of misalignment an MDC that was 
additionally integrated within $k_y=\pm\SI{0.15}{\angstrom^{-1}}$ around the 
$\Gamma$ point~\footnote{This corresponds to an angle of $\pm\SI{0.5}{\degree}$ 
in $k_y$ direction (denoted as tilt rotation in Figure~2 of 
Ref.~\cite{Strocov2014}) or a misalignment in the inplane rotation of 
$\pm\SI{5}{\degree}$.} is shown in Fig.~\refsubfig{fig:Cuts}{(c)}. 
The large difference between the two curves implies that the matrix 
element of the photoemission process is altered 
in presence of the EuS layer. 
The origin of this large change can be 
qualitatively 
understood by approximating the matrix element with the weights of the Fourier 
decomposition of the initial state wavefunction~\cite{Moser2017,Strocov2018}.
$\Gamma_0$ and $\Gamma_1$ correspond to the zeroth and first order in-plane 
Fourier coefficients, but the high $h\nu$ selects a higher order out-of-plane 
component from them. It seems plausible that  such weights of the higher 
harmonics in 
$k_z$  (i.e. sharp details of the wavefunction) may change in the presence of 
the EuS interface without causing a 
considerable change to the spectral lineshapes.
Therefore, we find clear evidence of a modification of 
the initial state wave function caused by the presence of the top layers.

\subsection{Local magnetic properties}

Zero-field measurements in the magnetic phase of bulk samples of EuS
 have shown that the local field at the muon
stopping position is on the order of
\SI{0.336}{T}~\cite{Eshchenko2009}.  In the weak transverse field
measurements that we report here, such strong magnetic fields will
cause the observed loss of $A_0$.  Therefore, our results are
consistent with previous measurements that reported ferromagnetic
ordering in the EuS thin layer~\cite{Wei2013,Katmis2016,Lee2016}.

The energy and temperature dependence of $A_0$ in 
Fig.~\ref{fig:EScans} can be
qualitatively understood, by comparing it to the simulated stopping
fractions:  
At high implantation
energies, the behavior can be fully explained by the temperature
dependence of muons in sapphire, which show an increases of $A_0$
towards low temperature~\cite{Prokscha2007,Krieger2017}.  The
pronounced loss of $A_0$ at low temperature and low implantation 
energy is
attributed to the magnetism in the EuS layer and at its interface.
The temperature independent full asymmetry at intermediate energies,
where most muons stop deep in the Bi$_2$Se$_3$ or Ti layer, is a clear
signature that any interface effects vanish further inside the
material.

We now turn to a quantitative estimation of the magnetic volume fraction in the 
samples.
From the calculated muon stopping profile of 
EuS/V$_{0.2}$(Bi,Sb)$_{1.8}$Te$_3$ 
we expect that $\sim\SI{23}{\percent}$ of the muons stop in the EuS
layer when using an implantation energy of \SI{1.5}{keV}. The measured
asymmetry is an ensemble average over all muons in the
sample. Therefore, if we assume that only muons stopping in the EuS
layer are depolarized, we expect  to observe a \SI{23}{\percent}
decrease relative to the full asymmetry in this sample as indicated 
with a bar in Fig.~\ref{fig:nAsy}.
This is consistent with our observation, since all muons 
stopping in the magnetic TI layer are already fully depolarized above 
\TCEuS$\sim\SI{16}{K}$. 
In the other samples, the EuS layer is slightly 
thinner~[Table~\ref{tab:thicknesses}] and following the same logic we 
expect a smaller drop of about \SI{15}{\percent} in the asymmetry. 
However, we observe a much larger drop instead.
This is a clear indication that large magnetic fields extend to regions outside 
the EuS layer, in particular into the interlayer beneath. The large
drop in $A_0$ can only be accounted for if muons stopping several nm
inside the Bi$_2$Se$_3$ and Ti layer are also depolarized.

\begin{figure*}[htb]
  \includegraphics[width=.9\linewidth]{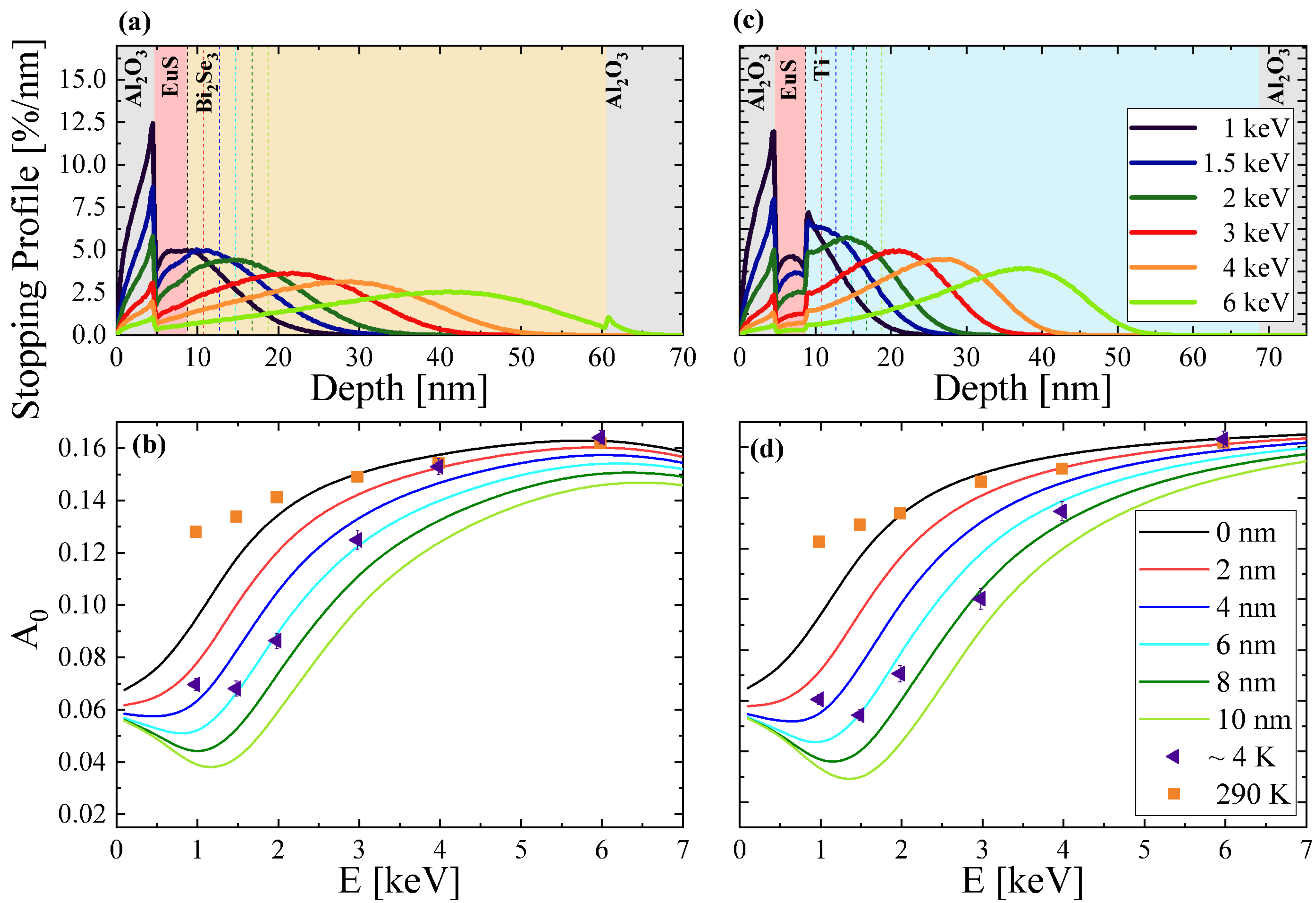}
  \caption{Muon implantation profiles for different implantation
    energies in the \textbf{(a)} 60\,QL~Bi$_2$Se$_3$ and \textbf{(c)}
    Ti samples. \textbf{(b)} and \textbf{(d)} show the corresponding
    initial asymmetries $A_0$ as a function of implantation energy $E$
    (i.e.~a zoom into
    Fig.~\refsubfig{fig:EScans}{(a)}~and~\refsubfig{fig:EScans}{(c)}). The solid
    lines show the calculated asymmetry where the muons stopping
    within the first $n$~nm of the interlayer are completely
    depolarized.  The depths of the magnetic proximity corresponds to
    the dashed lines of the same color in \textbf{(a)} and
    \textbf{(c)}.  }
\label{fig:profiles}
\end{figure*}

The fact that the size of the drop at \TCEuS\ in 
EuS/V$_{0.2}$(Bi,Sb)$_{1.8}$Te$_3$  is
correctly predicted by the simulation, further attests to the accuracy
of the \texttt{Trim.SP} results and justifies their use to make rough
estimation of the involved length scale of the region influenced by the
magnetic layer in the other samples. In these estimates we use the
calculated muon stopping profiles shown in
Figs.~\refsubfig{fig:profiles}{(a)}~and~\refsubfig{fig:profiles}{(c)}.
In order to evaluate the measured asymmetry, we assume that muons
stopping in EuS do not contribute, while those stopping in the
sapphire substrate contribute only \SI{42}{\percent} of their
polarization~\cite{Krieger2017}. In addition, muons stopping in the
interlayer are assumed to contribute fully to the polarization. The
result of this calculation is shown as a black line in
Figs.~\refsubfig{fig:profiles}{(b)}~and~\refsubfig{fig:profiles}{(d)}. Here, 
the total $A_0$ was scaled to match the point measured at RT and 
$E=\SI{6}{keV}$, where most muons stop in the interlayer. As expected, this 
curve overestimates $A_0$ at low implantation energy and recovers to the RT
values too quickly with increasing energy. To better account for our
measurements, we introduce an additional ``proximity magnetized''
layer of thickness $d$ in the near-interface region of the interlayer,
close to EuS. We assume that muons stopping in this layer are also
depolarized rapidly and do not contribute to the measured
asymmetry. The calculated curves for various values of $d$ are shown
in Fig.~\ref{fig:profiles}. They mimic more closely our measurements
for $d=4-\SI{8}{nm}$, though not perfectly. The discrepancy can, at
least partially, be attributed to our simplistic assumption of a
uniform, step-like magnetization profile, which is most probably not
the case in these samples.
Other possible sources of deviation include uncertainties in the
number of backscattered muons and effects of the magnetism in EuS onto
the Al$_2$O$_3$ capping. However, the results on 
EuS/V$_{0.2}$(Bi,Sb)$_{1.8}$Te$_3$ in
Fig.~\ref{fig:nAsy} indicate that the effect of the latter is very
small.

We conclude that our calculations provide a rough estimate of the
thickness of the affected region, between $4-\SI{8}{nm}$ for 
{\em both} 
EuS/(60\,QL)\,Bi$_2$Se$_3$ and EuS/(\SI{60}{nm})Ti, which is larger than the 
\SI{2}{nm} proximity that is typically
observed with PNR~\cite{Li2015,Katmis2016,Li2017}. This discrepancy is
primarily due to the higher sensitivity of \mSR\ to small magnetic
fields compared to PNR. Moreover, while an effective depolarization of
the muon spin can be caused by a strong field in an arbitrary 
direction, the PNR experiments are sensitive only to the
in-plane component of the magnetization. 
For example, in the EuS/V$_{0.2}$(Bi,Sb)$_{1.8}$Te$_3$ the local
magnetic fields are strong enough to completely depolarize the muons,
while the corresponding magnetic scattering length density in PNR is
very small~\cite{Krieger2017,Li2015}.
As discussed
in the \hyperref[sec:appendix]{Appendix}, the small negative shift of 
the field below \TCEuS\ could
be consistent with previously reported out-of-plane components of the
magnetism at the interface, generating long range stray fields that
would not have been seen with PNR~\cite{Wei2013,Lee2016}.

Note that the depolarization of the muons within $4-\SI{8}{nm}$ 
adjacent to the interface could be caused either by
proximity induced magnetism or by stray fields, e.g.~due to roughness
of the interface or finite magnetic domain size in the EuS layer. 
However, while the proximity effect, mediated by the TSS or bulk metallic 
states,  should occur close to the interface 
(within a few \si{\angstrom},~\cite{Menshov2013}), the 
relevant depth scale for stray fields is given by the length scale of the 
domains/modulation due to roughness~\cite{Tsymbal1994}.
In our samples the roughness is expected to be much smaller than 
\SI{4}{nm} and should be a minor contribution~\cite{Wei2013,Katmis2016}. 
Therefore, the observed depolarization several nanometers inside the 
interlayer is  most likely dominated by stray fields originating from magnetic 
domains. 

Surprisingly, there is a slow and gradual decrease of $A_0$ with
decreasing temperature in all samples already above the EuS transition
(indicated with a dashed line in Fig.~\ref{fig:nAsy}).  Such a
decrease is typically absent in non-magnetic samples and other undoped
TI thin films~\cite{Krieger2017}. For example, calibration
measurements on a gold film show a temperature independent initial
asymmetry, thus excluding experimental artifacts. Instead, this effect
could be a sign of interface magnetism persisting up to room
temperature, in agreement with Ref.~\cite{Katmis2016}. Note that Ti
has very small nuclear moments which are expected to produce only a
very slow damping of the \mSR\ asymmetry~\cite{Kossler1986,Amato2017}. 
It should thus be the ideal reference
sample as a topologically trivial metal. Therefore, the decrease of
the asymmetry with decreasing temperature, cannot be caused by the
presence of topological interface states.
There are two possible scenarios that could explain the observed decrease: 
First, it has an origin unrelated to interface magnetism. In this case, our 
results imply that there is no significant enhancement of the transition 
temperature in our samples of EuS/Bi$_2$Se$_3$. Second, the decrease is caused 
by a magnetic interface effect (within $\sim\SI{1}{nm}$ of the interface) that 
persists up to RT. However this would imply that the same 
effect is present in EuS/(\SI{60}{nm})Ti. 

Another unexpected feature in Fig.~\ref{fig:nAsy}, is that even below
\TCEuS, the curves measured in both samples (EuS/(60\,QL)\,Bi$_2$Se$_3$ and 
EuS/(\SI{60}{nm})Ti) are identical within our experimental accuracy. This is 
a strong indication that the magnetic fields extending into the
interlayer are very similar, but most importantly, they seem 
to be unaffected by the topology of the metallic states at the interface.
Since the size of this effect is the same in both materials, we conclude that
this property is intrinsic to the EuS/metal interface.

\section{Conclusion}\label{sec:Concl}

We combine several depth sensitive experimental techniques to
investigate the magnetic proximity effect in EuS/Bi$_2$Se$_3$. 
Our \mSR\ measurements reveal the presence of large local magnetic fields
that extend several nanometers away from the EuS layer and into the
adjacent non-magnetic layer. However, this length scale indicates that the main 
contribution to
the detected fields in the non-magnetic layer is stray fields from EuS
magnetic domains. A careful comparison between
EuS/Bi$_2$Se$_3$ and EuS/Ti reveals a qualitatively similar
behavior which implies that 
it does not rely  upon the presence
of topological states at the interface.
Rather, the 
dominant contribution to the observed local 
magnetic properties 
appears to be independent of the topology and the exact 
electronic structure at the interface.
Using anti-resonant SX-ARPES at the Eu M$_5$ pre-edge
we find that the dispersive electronic band structure of the buried 
Bi$_2$Se$_3$
layer remains mostly unaffected by the presence of the EuS and
Al$_2$O$_3$ layers.  
There is no clear signature of the previously predicted
interface states~\cite{Eremeev2015,Kim2017,Lee2014}, hinting at a 
different interface structure.  
However, we find a change of the relative spectral weight across different 
Brillouin zones, associated with an electronic reconstruction caused by the 
presence of EuS.

The combined \lem\ and SX-ARPES results show that there can be strong 
magnetic fields in the layer beneath EuS, unrelated to  topological interface 
states or the presence of strong magnetic exchange coupling. However, both of 
those are desirable when considering topological insulator/magnetic insulator 
interfaces for QAH devices.
Finally, to answer our initial question, the presented results can be fully 
explained without a need to introduce an interplay between topology and 
ferromagnetism at the EuS/Bi$_2$Se$_3$ interface.

\section*{Acknowledgments}
This work is based on experiments
performed at the Swiss Muon Source (S$\mu$S) and Swiss Light Source (SLS),
Paul Scherrer Institute, Villigen, Switzerland. 
The authors thank B.~P.~Tobler for his participation at the ARPES beamtime.
The work at PSI was supported by the Swiss National Science Foundation
(SNF-Grant No.~200021\_165910).
C.-Z.C. Y.-B.O and J.S.M. acknowledge the support from NSF grant no. 
DMR-1700137, Office of Naval Research (ONR) grant no. N00014-16-1-2657, and the 
Science and Technology Center for Integrated Quantum Materials under NSF grant 
no. DMR-1231319.
C.Z.C. thanks the support from Alfred P. Sloan Research Fellowship and ARO Young 
Investigator Program Award (W911NF1810198).

\appendix\section{Appendix: Detailed discussion of local fields measured with 
\lem}\label{sec:appendix}

\begin{figure}[htb]
  \includegraphics[width=0.95\linewidth]{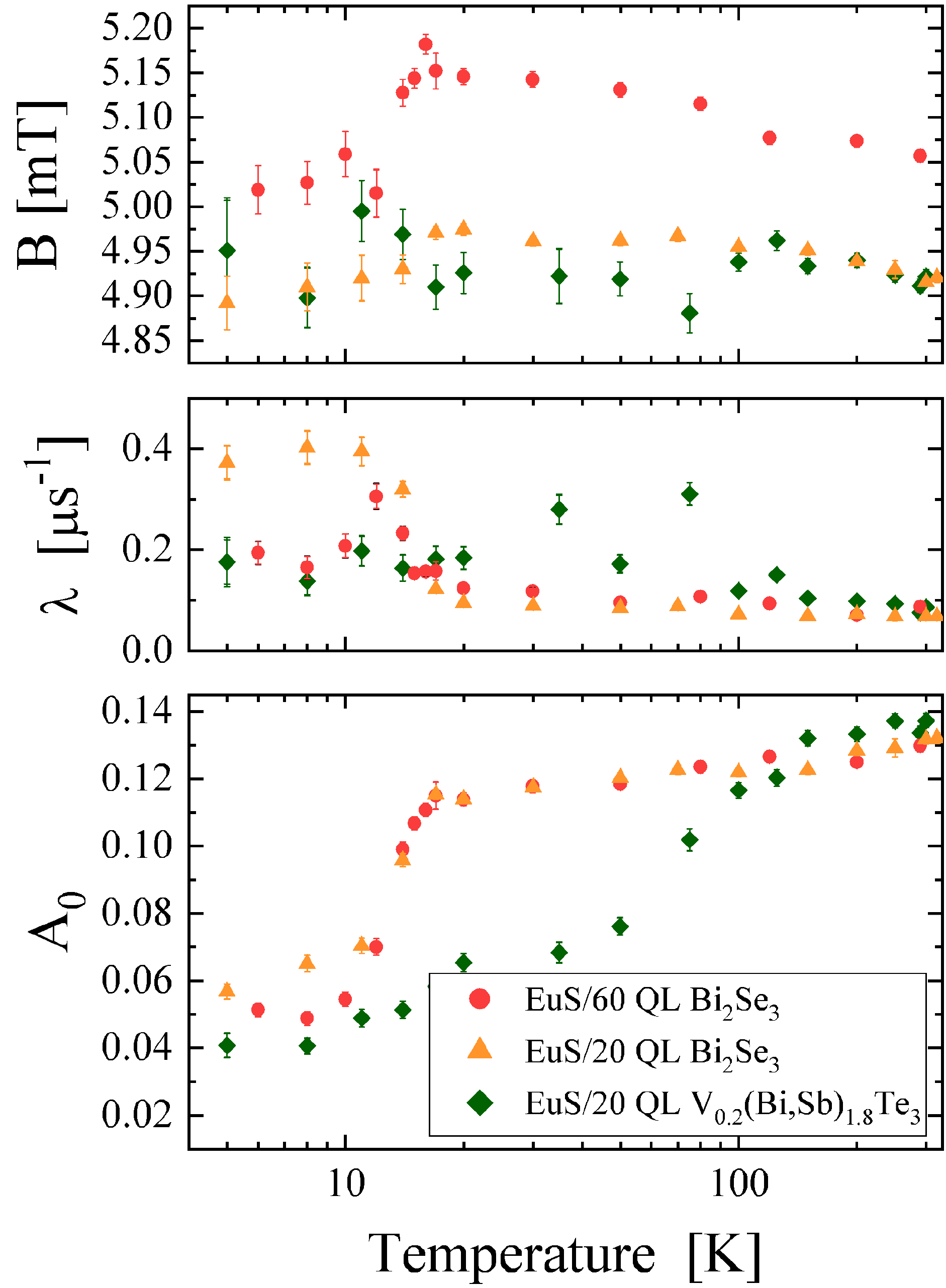}
\caption{Mean field, damping rate and initial asymmetry in wTF as a function of 
temperature for different TI samples, measured with an implantation energy 
of~\SI{1.5}{keV}. 
}
\label{fig:wTF}
\end{figure}

\begin{figure}[htb]
  \includegraphics[width=0.95\linewidth]{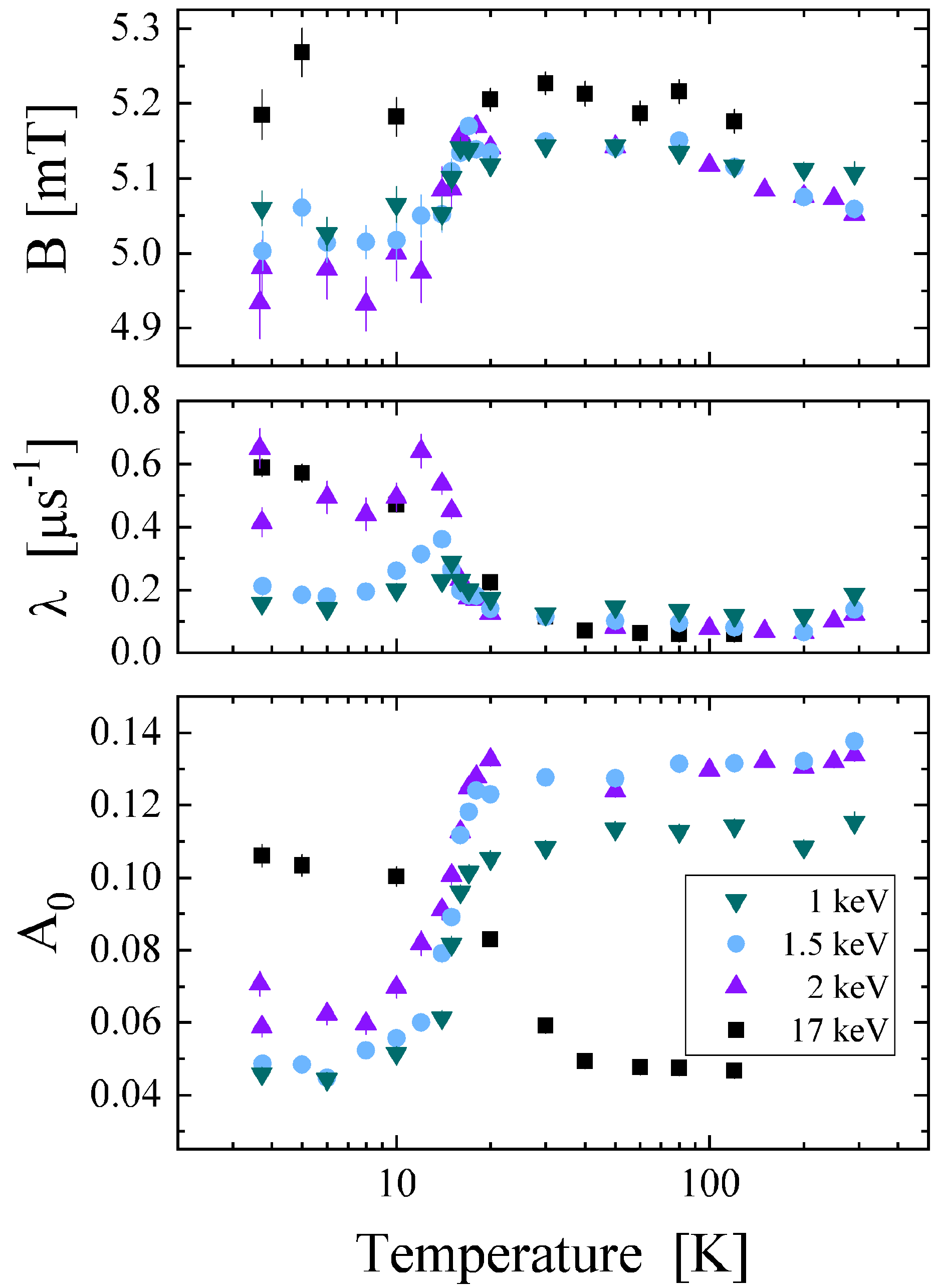}
\caption{Mean field, damping rate and initial asymmetry in wTF as a function of 
temperature in the EuS/(\SI{60}{nm})Ti sample at different implantation 
energies. The points at \SI{17}{keV} show the temperature dependence of the 
sapphire substrate.
}
\label{fig:Ti}
\end{figure}

The \mSR\ spectra were fitted to an exponentially damped cosine of the form
\begin{equation}\label{eq:AsyExp}
  A(t) = A_0e^{-\lambda t}\cos\left(\gamma_\mu Bt+ \varphi\right).
\end{equation}
In the main text we mainly discuss the initial asymmetry at 
$t= \SI{0}{\micro s}$, $A_0$. However, the damping rate $\lambda$ and the 
oscillation frequency $\omega=\gamma_\mu B$ are also affected by the magnetic 
transition. Here, $B$ is the  mean magnetic field at the
muons' stopping sites and $\gamma_\mu = 2 \pi \times \SI{135.5}{MHz/T}$  is the
muon gyromagnetic ratio.  Note that $B$ is mostly 
sensitive to 
out-of-plane component of the internal magnetic field~\footnote{In a 
transverse field ($B_{\mathrm{ext}}$)
the mean field 
\mbox{$B=\sqrt{{(B_{\mathrm{ext}}+B_\perp)}^2+{B_\parallel}^2}$}\mbox
{ $=\sqrt{{B_{ \mathrm { 
ext}}}^2+2B_{\mathrm{ext}}B_\perp+{B_{\mathrm{int}}}^2}$}
is more sensitive to 
out-of-plane components ($B_\perp$) than to in-plane components ($B_\parallel$) 
of the internal magnetic field $B_{\mathrm{int}}$.}. 
The initial phase $\varphi$ reflects the 
initial orientation of the implanted muons and depends also on the geometrical 
details of the spectrometer.
The temperature dependence of $B$, $\lambda$ and $A_0$ is shown in 
Fig.~\ref{fig:wTF} 
for the different topological insulator samples and in Fig.~\ref{fig:Ti} 
for different implantation energies in  
EuS/(\SI{60}{nm})Ti.

The damping rate $\lambda$ exhibits a peak at \TCEuS\ in some samples and 
remains larger than the RT value at low temperature.  This indicates an increase 
of the width of the static field distribution as well as some dynamic 
contributions at \TCEuS\ due to critical fluctuations. The mean field 
$B$ decreases at \TCEuS, except in EuS/V$_{0.2}$(Bi,Sb)$_{1.8}$Te$_3$.
The fact that there is no shift at low temperature in that 
sample~[Fig.~\ref{fig:wTF}] and no shift in EuS/(\SI{60}{nm})Ti at high 
implantation energies, implies that the shift is unlikely to be caused by a 
background contribution. Instead, it originates inside the samples, in 
particular from somewhere with no long range magnetic order, but still close 
to the interface region.
There are two interactions that may account for such a shift: stray fields 
and hyperfine coupling to polarized electrons that are screening the 
muon~\cite{yaouanc2011}.  The latter would require that a polarization of the 
conduction electrons is induced  several nm away from the interface. 
Moreover, the polarization would have to be out of 
plane unless the hyperfine coupling tensor had very large off-diagonal terms.
Therefore, it is more likely that stray fields are the source of the observed
field shift.
Note that in-plane dipolar fields will exhibit a symmetric field distribution 
of out-of-plane fields. 
This implies that a purely in-plane 
inhomogenity does not affect the out-of-plane mean field.
The observed shift of $B$ thus points to the presence of 
out-of-plane stray fields close to the interface in both EuS/Bi$_2$Se$_3$ and  
EuS/(\SI{60}{nm})Ti.

\end{document}